\documentstyle[prb,aps,psfig]{revtex}
\begin {document}
\bibliographystyle {plain}
\twocolumn[
\hsize\textwidth\columnwidth\hsize\csname @twocolumnfalse\endcsname 
\title
{\bf Localization of Spinwaves in the Quantum Hall Ferromagnet.}
 \author{A. G. Green\cite{Present}}
\address{Department of  Physics, Princeton University, New Jersey NJ 08544} 
\maketitle
\begin{abstract}
The quantum Hall ferromagnet at filling fraction $\nu =
1$ has some unusual properties due to the remarkable 
identity between the topological density of a 
spin distortion and the associated electrical charge 
density. We investigate the localization
 of spinwaves by coupling to a scalar disorder 
 potential via their topological density. A low energy 
 description of the system in terms 
of a non-linear sigma model of unitary supermatrices 
is derived. All states of this model are localized 
in two dimensions. A possible experimental signature 
of these effects in photo-luminescence is suggested.

\end{abstract}
\pacs{PACS numbers:73.40.Hm, 76.50.+g, 75.30.Ds}
]
The ground-state of a two-dimensional electron gas(2DEG) at exact filling of
a single Landau level (filling fraction $\nu=1$) is strongly ferromagnetic.
The properties of this quantum Hall ferromagnet(QHF) are profoundly affected
by the topological nature of the quantized Hall plateau. 
In a quantum Hall state, there is a commensuration between the magnetic
flux through the 2DEG and the electrical charge density. 
This leads to the identification of the topological charge density of a 
spin distortion  with an associated electrical charge density\cite{Sondhi}
(the Berry phase induced by a spin distortion may 
be reproduced by the Aharonov-Bohm phase induced by a fictitious magnetic flux).
As a consequence, the elementary
excitations formed as the filling fraction is moved slightly away from $\nu=1$
are electrically and topologically charged objects known as 
Skyrmions\cite{Sondhi,Skyrmion_expts}.
In addition, 
spinwaves may couple to a scalar disorder potential via their topological
density\cite{My_paper}.
The geometrical structure of this disorder interaction 
is very similar to minimal coupling of quantum particles to a random flux and 
the spinwave system provides a novel realization of this intently studied 
problem\cite{Aronov}.
In this paper, we investigate the localization 
of spinwaves in circumstances where the disorder potential is sufficiently weak
that the ground-state remains ferromagnetic and coupling to spinwaves is the
dominant effect. 
In the first section, we use diagrammatic techniques to demonstrate
a decoupling of 
charge-density fluctuations from fluctuations in exchange-energy.
In the second, we use supersymmetry\cite{Efetov} to 
construct a low energy description of the system in terms of a 
non-linear sigma model of unitary super-matrices. All states of this model
are localized in two dimensions\cite{Russians}. A possible
experimental signature of these effects  in photo-luminescence is suggested.

Our starting point is a spinwave expansion about the continuum field theory of 
the QHF proposed by Sondhi {\it et al}\cite{Sondhi,My_paper}.
The effective action and electrical current density for small fluctuations, 
${\bf l}=(l_1,l_2,0)$, about the ferromagnetic ground-state,  $\bar {\bf n}=(0,0,1)$,
 are given by
\begin{eqnarray}
S&=&
\int d^2x dt 
\left[
\frac{1}{2}  \bar l \left( \frac{\bar \rho}{2}
\partial_t - \rho_s \nabla^2- \bar \rho g B \right)  l
- J_0({\bf x}) U({\bf x})
\right],
\label{spinwave_action}
\\
J_{\mu}&=&
i\frac{e \nu}{8 \pi} \epsilon^{\mu \nu \lambda}
\partial_{\nu} \bar l \partial_{\lambda} l.
\label{top_density}
\end{eqnarray}
${\bf n}=(l_1,l_2,\sqrt{1-|{\bf l}|^2})$  is an O(3)-vector order parameter describing 
the local polarization of the quantum Hall system. The first 
part of Eq.(\ref{spinwave_action}) is the usual Schr\"odinger effective action 
for spinwaves in a continuum ferromagnet.  We use the complex notation, 
$l=l_1+il_2$, $\bar l=l_1-il_2$. $\bar \rho$ is the electron density 
($\bar \rho = \nu / 2 \pi l_B^2$, where $l_B$ is the magnetic length), $\rho_s$ 
is the spin stiffness and $g$ is the Zeeman coupling, into which we
have absorbed the electron spin and the Bohr magneton for ease of notation. 
The second term Eq.(\ref{spinwave_action}) is due to the 
identity of charge and topological charge embodied in 
Eq.(\ref{top_density}).  It describes the interaction of spinwaves with a scalar
disorder potential, $U({\bf x})$. The Coulomb self-interaction of the spinwave
charge density has been neglected, since it is higher order in both derivatives 
and in the spinwave field.
Throughout the bulk of the calculations
presented in this paper, we will ignore the Zeeman term in 
Eq.(\ref{spinwave_action}). It is a simple matter to re-introduce it at
the end of our calculations.

The correlations in the disorder potential felt by the two-dimensional
electron gas in GaAs heterostructures are conveniently modeled as 
follows:\cite{Efros}
\begin{equation}
\langle \langle U_{\bf q} U_{{\bf q}'}  \rangle \rangle 
=
(2\pi)^2 \delta({\bf q} + {\bf q}') 
\gamma' \frac{e^{-2|{\bf q}|d}}{|{\bf q}|^2}
.
\label{realistic_correlations}
\end{equation} 
$d$ is the width of the insulating spacer layer separating the electrons 
from the ionized donor impurities. $\gamma'$ is a measure of the disorder 
strength and is related to the area density of donor impurities, $n_d$; $\gamma'
= (e \sqrt{n_d}/2 \epsilon)^2(\nu/8 \pi)^2$. Our aim in 
this paper is to determine
the universal features that arise due to the geometrical structure of the
interaction. To simplify our explicit calculation, we
assume a Gaussian, $\delta$-function correlated distribution for the 
disorder potential:
\begin{eqnarray}
\langle \langle U_{\bf q} U_{{\bf q}'} \rangle \rangle 
&=&
(2\pi)^2 \gamma \delta({\bf q} + {\bf q}') 
\label{disorder_correlations}
\end{eqnarray}
We return to the more realistic correlations of Eq.(\ref{realistic_correlations})
towards the end of the paper.

Laid out in this way, the interaction of spinwaves with a disorder potential
has many similarities  to the problem of non-interacting Schr\"odinger particles
in a random flux\cite{Aronov}. To see this, one should integrate the interaction in 
Eq.(\ref{spinwave_action}) by parts. The result looks like minimal coupling
to a vector potential,
 $A_i=-(e \nu/16 \pi) \epsilon_{ij} \partial_j U({\bf x})$, 
aside from the absence of an $|{\bf A}|^2$ term (which leads to broken gauge 
symmetry in the present case).

The self-energy, calculated in the self-consistent Born approximation, with 
correlations in the disorder potential given by Eq.(\ref{disorder_correlations}),
is
\begin{eqnarray}
\Sigma^R({\bf p}, \omega)
&=&
\epsilon_{ij} \epsilon_{kl}
\begin{picture}(120,30)(0,20) 
\psfig{file=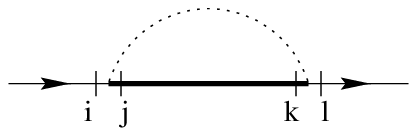}
\end{picture} 
\nonumber\\
\end{eqnarray}
In this diagram, full lines represent spinwave propogators, cross-bars indicate
 spatial derivatives of these propogators and the dotted lines represent disorder 
correlations. The thick line represents a full spinwave propogator including the 
self-energy; $G^{-1}=G_0^{-1} + \Sigma$. The real part of the self-energy 
may be absorbed into a renormalization of the spinwave stiffness. The 
imaginary part is given by
\begin{equation}
{\cal I}m \Sigma^R({\bf p}, \omega)
=
\gamma {\cal I}m
\int \frac{d^2q}{(2 \pi)^2}
({\bf p} \times {\bf q})^2 G^R({\bf q}, \omega).
\label{self_energy1}
\end{equation}
To lowest order in the disorder strength, this has the solution
\begin{equation}
{\cal I}m \Sigma^R({\bf p}, \omega)
=
\frac{\gamma}{8 \rho_s^2} \left( \frac{\bar \rho \omega}{2} \right)
|{\bf p}|^2.
\label{self_energy2}
\end{equation}
The disorder averaged spinwave Green's function is then
\begin{eqnarray}
\langle G^R({\bf p}, \omega) \rangle
&=&
\left(
\rho_s |{\bf p}|^2 -\frac{\bar \rho \omega}{2} 
+i\frac{\gamma}{8 \rho_s^2} \left( \frac{\bar \rho \omega}{2} \right)
|{\bf p}|^2
\right)^{-1}
\nonumber\\
&\approx&
\left(
\rho_s |{\bf p}|^2 -\frac{\bar \rho \omega}{2} 
+i\frac{\bar \rho}{2 \tau_{\omega}}
\right)^{-1}
\nonumber\\
\tau_{\omega}
&=&
\frac{4 \bar \rho \rho_s^3}{\gamma}
\left( \frac{\bar \rho \omega}{2} \right)^{-2}.
\end{eqnarray}
In writing down the final expression for the disorder averaged Green's function,
we have removed a factor of $\left(1+i \gamma \bar \rho \omega/16 \rho_s^3 \right)$
and Taylor expanded the denominator in powers of $\gamma \bar \rho \omega/16 \rho_s^3$.
Since this latter quantity is a small parameter in our perturbative expansion,
this Taylor expansion is justified. 

The most important point to notice here is that the scattering time diverges as
the frequency goes to zero. The spectral weight is concentrated in
an energy range $1/2 \tau_{\omega}$ of the bare pole at $\bar \rho \omega /2
= \rho_s |{\bf q}|^2$, such that $1/2 \tau_{\omega} \ll 
\bar \rho \omega /2$. This should be compared to the case of electrons scattering
from a random scalar potential\cite{Efetov}, where the bare pole
of the Green's function is near to the Fermi energy, $E_F$, and the scattering
rate, $1/\tau$, is constant. The validity of the perturbative expansion for
electrons depends upon the smallness of the parameter $1/(\tau E_F)$. In the 
present case, the extra derivatives in the interaction vertex lead to the 
existence of a small expansion parameter without the existence of a Fermi surface.
From a calculational point of view, $\bar \rho \omega/2$ plays the role of a 
frequency dependent
Fermi surface and momentum integrals may be carried out in the same way as one 
would in the case of electrons.

\section{Diffusons and Cooperons}
The diffusive and weak localization effects in a disordered system are 
determined by the diffuson
\begin{eqnarray}
\lefteqn{
D({\bf p},{\bf q}, {\bf k}, \Omega, \Omega+\omega)
=
\left.
\begin{picture}(135,35)(0,30) 
\psfig{file=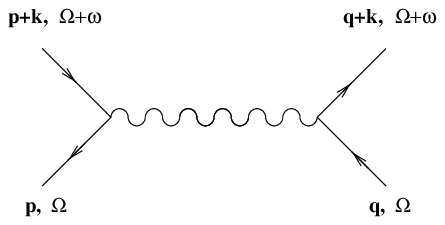}
\end{picture}
\right.
}
\nonumber\\
&=&
\epsilon_{ij} \epsilon_{kl}
\left.
\begin{picture}(170,60)(0,25) 
\psfig{file=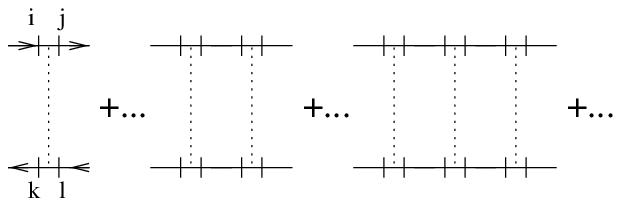}
\end{picture}
\right.
\nonumber\\
\nonumber\\
\label{Diffusons}
\end{eqnarray}
and Cooperon modes
\begin{eqnarray}
\lefteqn{
C({\bf p},{\bf q}, {\bf k}, \Omega, \Omega+\omega)
=
\left.
\begin{picture}(135,35)(0,30) 
\psfig{file=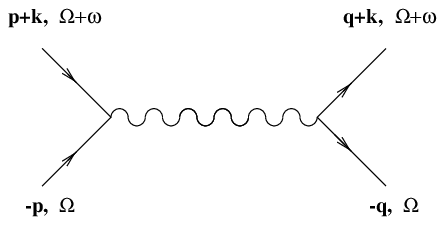}
\end{picture}
\right.
}
\nonumber\\
&=&
\epsilon_{ij} \epsilon_{kl}
\left.
\begin{picture}(170,60)(0,25) 
\psfig{file=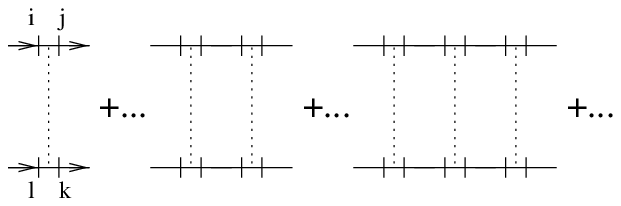}
\end{picture}
\right.
\nonumber\\
\nonumber\\
\label{Cooperons}
\end{eqnarray}
respectively. Notice that the interaction vertex is anti-symmetric under time
reversal and, therefore, occurs with the opposite sign in the ladder summation
for the Cooperon, compared with the Diffuson. The ladder approximation to these 
diagrams, as indicated in 
Eqs.(\ref{Diffusons},\ref{Cooperons}), may be summed to give two Dyson's equations
that may be represented by the following diagram: 
\begin{equation}
\begin{picture}(220,35)(0,5) 
\psfig{file=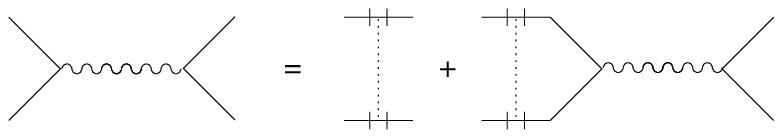}
\end{picture} 
\label{Dyson}
\end{equation}
In fact, these equations are rather difficult to handle. The trouble is that
the geometrical factors, involving cross-products of momenta, are not easy to
factorize. One way to negotiate these difficulties is to write the interaction
vertex as the sum of four vertices;
\begin{equation}
\epsilon_{ij} \epsilon_{kl}
\begin{picture}(200,35)(0,25) 
\psfig{file=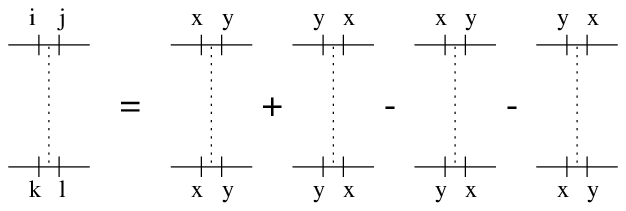}
\end{picture} 
\end{equation}
The diffuson and Cooperon diagrams are decomposed similarly into the sum of
diagrams;
\onecolumn
\begin{equation}
D=D_{xxxx}+D_{xxxy}+D_{xxyx}+ \hbox{13 other terms},
\end{equation}
where the suffices label the derivatives on the external legs of the contributing
diagrams. We wish to evaluate these diagrams in the limit of small momentum,
${\bf k}$, and frequency, $\omega$. Consider the momentum integral over some
internal rung of a ladder diagram when ${\bf k}=0$;
\begin{equation}
\begin{picture}(150,35)(0,22) 
\psfig{file=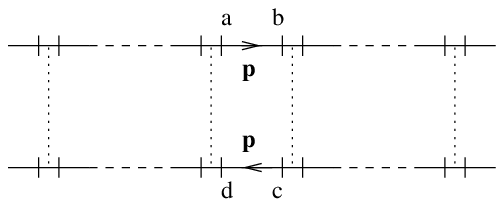}
\end{picture} 
=
... \int \frac{d^2p}{(2 \pi)^2} 
p_a p_b p_c p_d 
{\cal G}({\bf p}, \Omega+\omega)
{\cal G}({\bf p}, \Omega)...
\end{equation}
\newline
The angular integration gives zero if $a \ne b=c=d$ (or a permutation of this).
The remaining non-zero rungs are
\begin{equation}
\begin{picture}(220,120)(0,0) 
\psfig{file=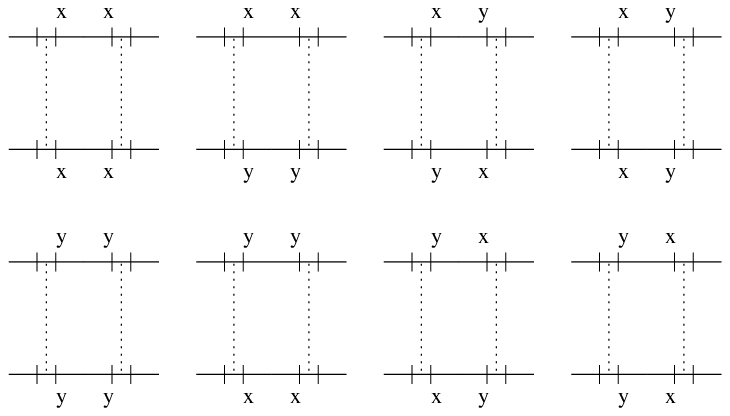}
\end{picture} 
\end{equation}
Only elements of $D$ corresponding to these rungs are non-zero. In addition, 
rotational symmetry requires that 
\begin{eqnarray}
& &D_{xxxx}=D_{yyyy}, 
\;\;\;\;
D_{xxyy}=D_{yyxx}, 
\nonumber\\
& &D_{xyxy}=D_{yxyx}, 
\;\;\;\;
D_{xyyx}=D_{yxxy}. 
\end{eqnarray}
Using these symmetries in Eq.(\ref{Dyson}), with ${\bf k}=0$, the Dyson's equation 
decouples into two pairs of coupled equations,
\begin{equation}
\begin{picture}(350,130)(0,0) 
\psfig{file=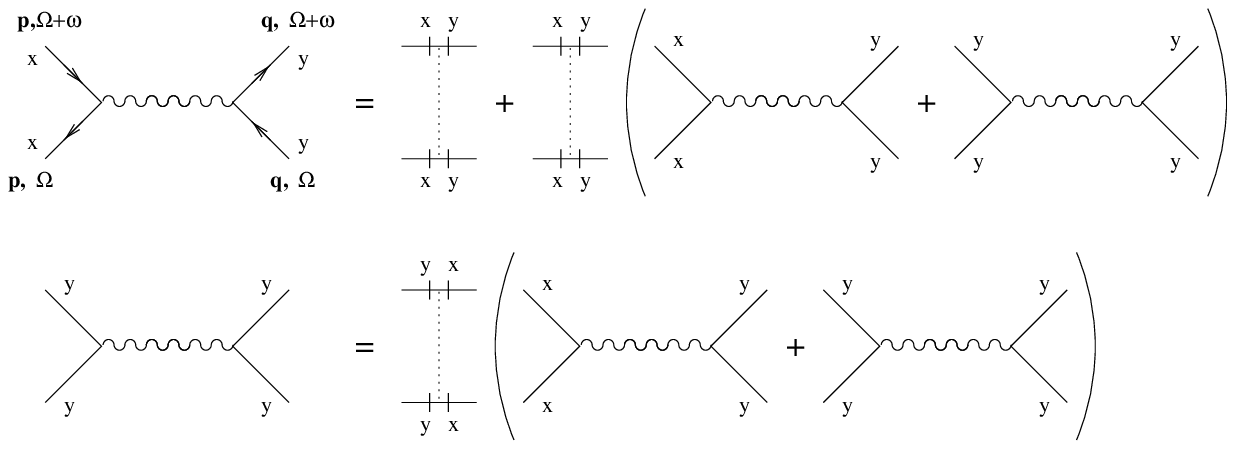}
\end{picture}
\label{massless_Dyson} 
\end{equation}
relating $D_{yyyy}$ to $D_{xxyy}$ and 
\begin{equation}
\begin{picture}(350,130)(0,0) 
\psfig{file=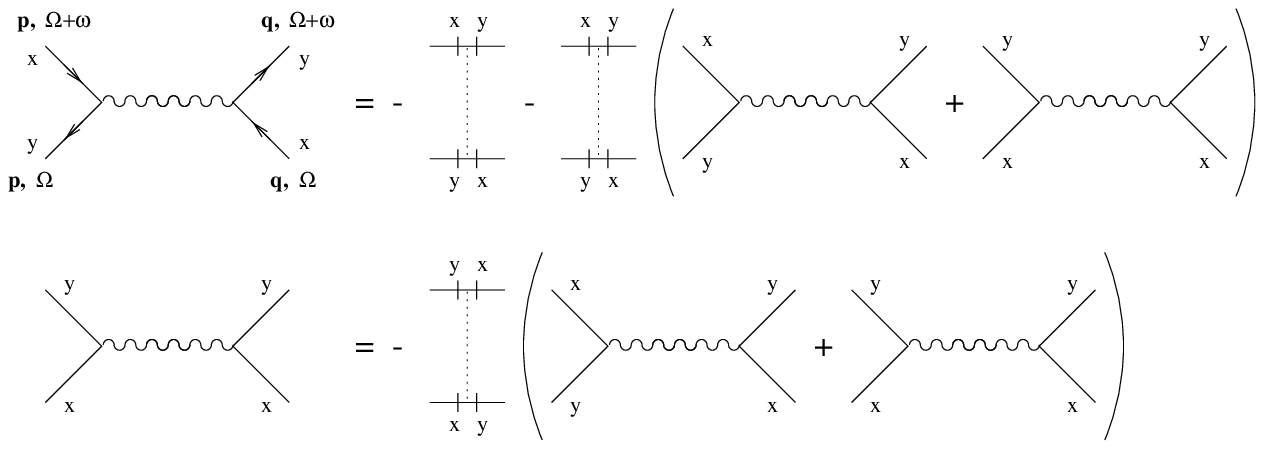}
\end{picture} 
\label{massive_Dyson}
\end{equation}
\twocolumn
relating $D_{xyyx}$ to $D_{yxyx}$. 
The components, $D_{xyxy}$, {\it etc.} of $D$, describe the propogation of 
charge-density fluctuations (the fields on the left-most legs occur in the 
combinations $\partial_1 \bar l \partial_2 l$ and 
$\partial_2 \bar l \partial_1 l$, the charge-density being proportional to 
$\epsilon_{ij}\partial_i \bar l \partial_j l$) and the components $D_{xxxx}$, 
{\it etc.} describe the propogation of exchange-energy fluctuations (the fields 
on the external legs occur in the combinations 
$\partial_1 \bar l \partial_1 l$ and
$\partial_2 \bar l \partial_2 l$, the exchange-energy being proportional to 
$\partial_i \bar l \partial_i l$).
 The decoupling of the 
Dyson's equation, therefore, amounts to decoupling of charge and energy 
propogation. An alternative way of separating a Dyson's equation such as 
Eq.(\ref{Dyson}) is presented in Ref.[\onlinecite{Bhatt}].
Written in the above manner,
the physical eignemodes of the scattering process are apparent. 
Using the ansatz
\begin{equation}
D_{abcd}({\bf p},{\bf q}, {\bf k}, \Omega,\omega)
=(p+k)_ap_b(q+k)_cq_d \tilde 
D_{abcd}({\bf k}, \omega, \Omega)
\end{equation}
Eq.(\ref{massless_Dyson}) reduces to
\begin{eqnarray}
\tilde D_{xxyy}
&=&
\gamma
+
\left[
\tilde D_{xxyy} \langle \cos^2 \theta \sin^2 \theta \rangle_{\theta}
+
\tilde D_{yyyy} \langle \cos^4 \theta \rangle_{\theta}
\right]
\nonumber\\ 
\;\;\;\;\;\;\; & & \times
\gamma
\int \frac{d^2q}{(2 \pi)^2} |{\bf q}|^4 
 G^A({\bf q},\Omega) G^R({\bf q}, \Omega+\omega),
\nonumber\\
\tilde D_{yyyy}
&=&
\left[
\tilde D_{xxyy} \langle \cos^4 \theta \rangle_{\theta}
+
\tilde D_{yyyy} \langle \cos^2 \theta \sin^2 \theta \rangle_{\theta}
\right]
\nonumber\\ 
\;\;\;\;\;\;\; & & \times
\gamma
\int \frac{d^2q}{(2 \pi)^2} |{\bf q}|^4 
 G^A({\bf q},\Omega) G^R({\bf q}, \Omega+\omega)
\end{eqnarray}
Evaluating the momentum integrals  and 
performing the angular averages, one obtains
\begin{eqnarray}
\tilde D_{xxyy}
\left( 1-i \frac{2 \omega \tau_{\Omega}}{3} \right)
&=&
\frac{4\gamma}{3}
+
\tilde D_{yyyy}
\left( 1+i 2  \omega \tau_{\Omega} \right)
\nonumber\\
\tilde D_{yyyy}
\left( 1-i \frac{2  \omega \tau_{\Omega}}{3} \right)
&=&
\tilde D_{xxyy}
\left( 1+i 2  \omega \tau_{\Omega} \right).
\end{eqnarray}
To lowest order in $\omega$, these equations have the solution
\begin{equation}
\tilde D_{xxyy}({\bf k}=0,\omega,\Omega)=
\tilde D_{yyyy}({\bf k}=0,\omega,\Omega)
=\frac{\gamma \bar \rho}{8 \tau_{\Omega}} \left( i \frac{\bar \rho \omega}{2} \right)^{-1}.
\end{equation}
$\tilde D_{xxyy}$ and $\tilde D_{yyyy}$, therefore, describe massless diffusive
modes. A similar calculation for $\tilde D_{xyyx}$ and $\tilde D_{yxyx}$ produces
a result that is finite as $\omega \rightarrow 0$. These correspond to massive 
modes. Expanding the Dyson's equation to lowest order in 
${\bf k}$, one obtains the result
\begin{equation}
\tilde D_{xxyy}({\bf k},\omega,\Omega)=
\tilde D_{yyyy}({\bf k},\omega,\Omega)=
\tilde D({\bf k},\omega,\Omega),
\end{equation}
where
\begin{equation}
\tilde D({\bf k},\omega,\Omega)=
\frac{\gamma \bar \rho}{8 \tau_{\Omega}} 
\left( i \frac{\bar \rho \omega}{2}  +D_o |{\bf k}|^2 \right)^{-1},
\end{equation}
and $D_0= \rho_s \Omega \tau_{\Omega}$ is the classical diffusion 
coefficient for the spinwaves. In principle, one should worry that, when ${\bf k}
\ne 0$, the Dyson's equations no longer decouple into two pairs of equations.
This may lead to a shift in the diffusion constant, $D_0$\cite{Aronov,Bhatt}. We
 find that, in this case, the result is unaffected to lowest order.
Finally, summing all of the massless contributions to the diffuson, we find
\begin{equation}
D({\bf p},{\bf q}, {\bf k}, \Omega, \Omega+\omega)
=
({\bf p}+ {\bf k}). {\bf p}({\bf q}+ {\bf k}). {\bf q}
\tilde D({\bf k},\omega,\Omega),
\label{Diffuson_prop}
\end{equation}
which is the massless propogator for exchange energy fluctuations. 
The Cooperon diagrams may be evaluated similarly. They are 
finite in the limit ${\bf k}, \omega \rightarrow 0$ and, therefore, massive. This is 
a consequence of the time reversal anti-symmetry. There is no contribution to 
weak localization from these modes.

\section{Supersymmetric Solution.}
We now develop a low energy theory for the interaction of
spinwaves with a weak disorder potential, using supersymmetric techniques.
The main subtleties of the current problem are
 in the handling of the geometrical factors in the interaction.
We will concentrate upon these features and refer the reader to the 
literature\cite{Efetov} for details of the super-symmetry itself.
As noted previously, this problem is very similar to that of
non-interacting particles in a random flux. An alternative derivation of the 
results presented here may be made using the techniques of
 Ref.[\onlinecite{Aronov}].

We wish to determine the disorder averages of dynamical quantities involving
$\langle G^A G^R \rangle$ and, therefore, introduce an eight component 
superfield,
\begin{equation}
\psi=
\left(
l^A \bar l^A l^R \bar l^R \chi^A \bar \chi^A \chi^R \bar \chi^R
\right),
\end{equation}
in the usual way\cite{Efetov}. The superscripts, $A/R$, 
label advanced and retarded sectors, the fields $\chi$ are anti-commuting and 
the overbar indicates complex conjugation. The presence of the commuting
and anti-commuting fields in $\psi$ removes the need to write the partition 
function explicitly in the denominator of correlation functions and allows the 
disorder average to be performed immediately. The resulting Lagrangian for the 
superfield, $\psi$, is
\begin{equation}
{\cal L}
=
\int 
\left[
-i \bar \psi \left(
-\rho_s \nabla^2 -\frac{\bar \rho}{2} \tilde \epsilon \right)
\psi
+ \gamma \left( \epsilon_{ij} \partial_i \bar \psi \partial_j \psi \right)^2
\right]
d^2r,
\end{equation}
where $\tilde \epsilon=(\Omega+\omega/2){\bf 1} +(\omega/2 +i \delta)\Lambda$
and $\Lambda$ is the diagonal supermatrix with elements $\pm 1$ in the 
advanced/retarded sectors. 
As is usual in the derivation of the supersymmetric sigma model, we assume
that $\omega \ll \Omega$. Disorder averaging may only induce cross correlations
between the Green's functions if the difference in frequencies is less than the 
scattering rate $\bar \rho \omega/2 \le 1/2 \tau_{\Omega}, 
1/2 \tau_{\Omega+\omega} \ll \bar \rho \Omega/2$, therefore this approximation
is justified.
 Next, we decouple the quartic interaction, introduced
by the disorder average, with a supermatrix field, $Q_{ij}$, where $i,j \in \{
x,y\}$ and each element of the $2\times2$ matrix $Q_{ij}$ is an $8\times8$ 
supermatrix;
\begin{eqnarray}
& &
\exp \left( -{\cal L}_{int}[\psi] \right)
=
\int \exp \left( -{\cal L}_{int}[\psi,Q_{ij}] \right)
DQ_{ij},
\nonumber\\
& &
{\cal L}_{int}[\psi,Q_{ij}]
\nonumber\\
&=&
\frac{\bar \rho}{2 \tau_{\Omega}} \int 
\hbox{Str}
\left(
\left(\frac{\bar \rho \Omega}{2 \rho_s} \right)^{-1}
\epsilon_{ki}
\partial_k \bar \psi Q_{ij} \partial_j \psi 
+\frac{1}{32 \rho_s}  Q_{ij}^2 
\right) d^2r
\end{eqnarray}
The supertrace, $\hbox{Str}$, is as defined in Ref.[\onlinecite{Efetov}] and 
summation over repeated spatial indices is implied. Integrating out the 
superfield, $\psi$, we obtain the following free energy functional for $Q_{ij}$:
\begin{equation}
F[Q_{ij}]=
\int \hbox{Str} \left[ -\frac{1}{2}  ( \ln G^{-1} )
+ \frac{\bar \rho}{64 \tau_{\Omega} \rho_s}Q_{ij}^2 \right] d^2r, 
\label{free_energy_functional}
\end{equation}
where $G({\bf r}, {\bf r}', Q)$ is the supermatrix Green's function of the field 
$\psi$ and satisfies the equation
\begin{eqnarray}
& &
\left(
-\rho_s \nabla^2 -\frac{\bar \rho}{2} \tilde \epsilon
+i \frac{\bar \rho}{2 \tau_{\Omega}}
\left(\frac{\bar \rho \Omega}{2 \rho_s} \right)^{-1}
\epsilon_{ki} 
\left[ Q_{ij} \partial_j \partial_k+ \partial_k Q_{ij} \partial_j \right]
\right)
\nonumber\\
& & \;\;\;\;\;\;\;\;\;\;\;\;\;\;\;\;\;\;\;\;\;\;\;\;\;\;\times
G({\bf r}, {\bf r}', Q)
=
i \delta({\bf r}-{\bf r}').
\label{Greens_function_eqn}
\end{eqnarray}
The saddle point equation for $Q_{ij}$ is
\begin{equation}
Q_{ij}=
\frac{8 \rho_s}{\gamma} 
\left(\frac{\bar \rho \Omega}{2 \rho_s} \right)^{-1}
\gamma \int \frac{d^2p}{(2 \pi)^2}
\epsilon_{ki}p_k p_j G({\bf p}, Q_{ij} ).
\end{equation}
This is precisely the self-consistency equation that was solved previously to find
 ${\cal I}m \Sigma$. The solution is
\begin{equation}
Q_{ij}=\epsilon_{ij} V \Lambda \bar V,
\end{equation}
where $V$ is an arbitrary, unitary supermatrix such that $V \bar V=1$. The 
diagonal terms of $Q_{ij}$ are zero at the saddle point and, 
therefore, correspond to massive modes. The off-diagonal components, however,
sit in a Mexican hat potential and have massive longitudinal fluctuations
and massless transverse fluctuations. At the saddle point, 
$Q_{ij}\sim \epsilon_{ki} \partial_k \bar \psi \partial_j \psi$, and, therefore,
the diagonal elements of $Q_{ij}$ describe charge density fluctuations,
$Q_{ii} \sim \epsilon_{ij} \partial_i \bar \psi \partial_j \psi$ and the 
off-diagonal elements describe exchange energy fluctuations, $
\epsilon_{ij} Q_{ij} \sim \partial_i \bar \psi \partial_i \psi$.
This is the same decomposition into massive and massless modes as was found
diagrammatically in the preceding section.  Let us define
\begin{eqnarray}
Q_{ij}&=&\epsilon_{ij} Q_s+\hbox{diag} (Q_{c1},Q_{c2})
\nonumber\\
Q_s&=&\frac{1}{2} \epsilon_{ij} Q_{ij}
\nonumber\\
Q_{c1}&=& Q_{11}
\nonumber\\
Q_{c2}&=& Q_{22}
\label{Q}
\end{eqnarray}
At the saddle point $Q_s=V\Lambda \bar V$ and $Q_c=0$. 
Expanding the free energy functional, Eq.(\ref{free_energy_functional}), to quadratic order in 
fluctuations of $Q_{ij}$ about the saddle point, we obtain
\begin{eqnarray}
S&=&
\frac{\bar \rho}{64 \tau_{\Omega} \rho_s}
\int d^2r \hbox{Str}
\left[
 Q_{ij}^2
\right]
\nonumber\\
& &
-\frac{\bar \rho}{8 \tau_{\Omega}^2} 
\left( \frac{\bar \rho \Omega}{2 \rho_s} \right)^{-2}
\int \frac{d^2p}{(2 \pi)^2}
\int \frac{d^2q}{(2 \pi)^2}
\nonumber\\
& & \times
\hbox{Str}
\left[
\epsilon_{ki} p_j (p+q)_k G({\bf p}) \delta Q_{ij}({\bf q})
\right.
\nonumber\\
& &
\;\;\;\;\;\;\;\times \left.
\epsilon_{nl} (p+q)_m p_n G({\bf p}+{\bf q}) \delta Q_{lm}(-{\bf q})
\right].
\end{eqnarray}
The ${\bf p}$-integral is only finite for the products $G^AG^R$, which
are generated by the components of $Q$ off-diagonal in the advanced and retarded sectors. Henceforth, we use $Q$ to indicate supermatrices with only these off-diagonal components non-zero. Carrying out these
integrals and substituting $Q_s$, $Q_{c1}$ and $Q_{c2}$ from Eq.(\ref{Q}), we find the following effective action for $\delta Q_s$:
\begin{eqnarray}
 S&=&
\frac{1}{32 \rho_s}
\int \frac{d^2q}{(2 \pi)^2}
\left(-i \bar \rho \omega/2 + D_0 |{\bf q}|^2 \right)
\nonumber\\
& &
\;\;\;\;\;\;\;\;\;\;\;\;\;\;\;\;\;\;\;\;\;\;\;\;\;\;\;\;\;\times
\hbox{Str}
 \left[ \delta Q_s({\bf q})\delta Q_s(-{\bf q}) \right],
\label{delta_S}
\end{eqnarray}
where
\begin{equation}
D_0
=
  \rho_s \tau_{\Omega}\Omega.
  \label{D}
\end{equation}
The corresponding actions for $\delta Q_{c1}$ and $\delta Q_{c2}$ contain massive
propogators. It is important that one should not
simply ignore these massive modes. They may lead to a renormalization of the 
diffusion constant for the massless modes\cite{Bhatt,Aronov}. Here, we find that 
this is not the case. The cross-terms between $\delta Q_s$ and $\delta Q_{c1}
/\delta Q_{c2}$ are proportional to $|{\bf q}|^2$. Integrating out the massive modes 
induces a $|{\bf q}|^4$-term in the $\delta Q_s$ propogator, which we neglect at small
momentum. 
Keeping only fluctuations of $Q$ over the saddle point manifold, Eq.(\ref{delta_S})
 determines the coefficients of the terms in the sigma model;
\begin{eqnarray}
F[Q_s]
&=&
\frac{1}{128\rho_s}
\hbox{Str}
\int \left[
D_0 |\nabla Q_s|^2 -2i \left(\frac{\bar \rho \omega}{2} \right) \Lambda Q_s 
\right]
d^2r
\label{sigma_model}
\end{eqnarray}
$D_0$ is the classical diffusion constant given by Eq.(\ref{D}). It may be written in the form
$D_0=\bar \rho \tau_{\Omega} v_{\Omega}^2/2$, where $v_{\Omega}^2=(dE/d{\bf p} )^2\left.
\right|_{E=\Omega/2}=2 \rho_s \Omega/\bar  \rho$. This energy 
functional describes the diffusive
propogation of exchange energy density fluctuations. All states are 
localized in this model\cite{Efetov,Russians}, with a localization length
\begin{equation}
\xi=v_{\Omega} \tau_{\Omega} \exp \left[ \frac{\pi^2 D_0^2}{64 \rho_s^2} \right]
\sim 
\Omega^{-3/2} \exp \left[ \Omega^{-2} \right],
\end{equation}
which is divergent in the $\Omega$, ${\bf q}=0$ limit. This divergence is a 
natural
consequence of the existence of an SU(2) global symmetry (broken to U(1)
with the inclusion of the Zeeman coupling). 
Inclusion of the Zeeman coupling modifies the parameters of the sigma model 
such that $\Omega$ is replaced by 
$\Omega-gB$. There are no spinwave states below the Zeeman gap and the 
localization length diverges as the frequency approaches this, ${\bf q}=0$,
limit. Realistic correlations in the disorder potential, given by 
Eq.(\ref{realistic_correlations}), lead to a further modification of the 
parameters in the sigma model. The scattering time is given by\cite{My_paper}
\begin{equation}
\tau_{\Omega}'=\frac{4 \bar \rho \rho_s^2}{\gamma'}
\left( \frac{\bar \rho \Omega}{2} \right)^{-1},
\end{equation}
which again diverges as $\Omega \rightarrow 0$ such that $1/\tau_{\Omega} \ll \bar \rho \Omega /2$.
Subsequent calculations follow through as before, with some additional complications.
Finally, one obtains the same supersymmetric sigma model, Eq.(\ref{sigma_model}), with
\begin{equation}
D_0'=\rho_s  \tau_{\Omega}'\Omega=\frac{8 \rho_s^3}{\gamma'}.
\end{equation}
The diffusion constant is now independent of $\Omega$ and
the localization length no longer has an exponential dependence upon frequency, 
but a power law dependence;
\begin{equation}
\xi' \propto l_B \left(\frac{\Omega}{\rho_s} \right)^{-1/2} .
\end{equation}
The numerical prefactor in this expression has an exponential dependence upon 
the disorder strength. Estimates assuming a donor density, $n_d$, of the same
order as the electron density, $\bar \rho$, gives prefactors upwards of $10^3$.
Experimental determination of the diffusion coefficient and localization length
may be possible using space/time-resolved photo-luminescence. In the 
QHF, magneto-excitons and spinwaves are identical\cite{Kallin}. If, in addition,
the sub-band wavefunction of valence holes is centered in the same position as 
the wavefunction of electrons in the two-dimensional electron gas, then excitons
have the same Hamiltonian as magneto-excitons\cite{Nigel}. 
Localization and diffusion of excitons in other systems have 
been measured using photo-emission spectroscopy\cite{photoemission}. The same 
techniques may be applicable here.

Above a threshold disorder strength, the groundstate ceases to be ferromagnetic
and contains a random distribution of Skyrmions and anti-Skyrmions\cite{My_paper,Nazarov}.
 In certain circumstances,
the main effect of this topologically non-trivial groundstate upon the spinwave action is to 
introduce an effective magnetic field and mass term\cite{Polyakov}. Both the magnetic flux and 
the mass are
proportional to the topological density of the background. This is yet another variant of the 
random flux problem. In this case, the effective random flux may be very large 
and localization may have further experimental consequences\cite{not_done_yet}.

In summary, we have considered the effect of weak disorder upon the spinwave 
dynamics in the QHF.
We find a spinwave scattering time that diverges as frequency approaches zero,
thus ensuring the validity of perturbative and supersymmetric techniques 
despite the lack of a Fermi-surface. 
Charge and exchange-energy  fluctuations are found to decouple, the former 
being massive and the latter diffusive. The low energy dynamics of this system are 
described by a non-linear sigma model of unitary supermatrices. All states in this 
model are localized. A possible signature of these effects may be found in 
space/time-resolved photoemission spectroscopy.
\newline
\newline
Thanks to Nigel Cooper, Michael Fogler, Duncan Haldane, 
Matt Hastings, Shivaji Sondhi  and Alexei Tsvelik for comments and suggestions.

\end{document}